\def\year{2019}\relax
\documentclass[letterpaper]{article}
\usepackage{aaai19}
\usepackage{times}
\usepackage{helvet}
\usepackage{courier}
\usepackage{amsmath}
\usepackage{amssymb}
\usepackage{multirow}
\usepackage{graphicx}
\usepackage{subcaption}
\usepackage{array,booktabs}
\usepackage{url}
\usepackage{gensymb}
\usepackage{lipsum}% http://ctan.org/pkg/lipsum

\newcommand{\Z}{\mathbf{Z}}
\newcommand{\x}{\mathbf{x}}
\newcommand{\xhat}{\hat{\mathbf{x}}}

\newcommand{\Zi}{\mathbf{Z}^{-i}}
\usepackage{theorem}
\newcommand{\qed}{\rule{1.5ex}{1.5ex}}  % end of proof

\newtheorem{lemma}{Lemma}

\begin{document}
% The file aaai.sty is the style file for AAAI Press 
% proceedings, working notes, and technical reports.
%
\title{DoPAMINE:\\ Double-sided Masked CNN for Pixel Adaptive Multiplicative Noise Despeckling}

\author{
Sunghwan Joo\textsuperscript{\rm 1},
Sungmin Cha\textsuperscript{\rm 1},
Taesup Moon\textsuperscript{\rm 1}\\
\textsuperscript{\rm 1} Department of Electrical and Computer Engineering, Sungkyunkwan University\\ \{shjoo840, csm9493, tsmoon\}@skku.edu\\
}

\maketitle
\begin{abstract}
\begin{quote}
We propose DoPAMINE, a new neural network based multiplicative noise despeckling algorithm. Our algorithm is inspired by Neural AIDE (N-AIDE), which is a recently proposed neural adaptive image denoiser. While the original N-AIDE was designed for the additive noise case, we show that the same framework, \emph{i.e.}, adaptively learning a network for pixel-wise affine denoisers by minimizing an unbiased estimate of MSE, can be applied to the multiplicative noise case as well. Moreover, we derive a double-sided masked CNN architecture which can control the variance of the activation values in each layer and converge fast to high denoising performance during supervised training. In the experimental results, we show our DoPAMINE possesses high adaptivity via fine-tuning the network parameters based on the given noisy image and achieves significantly better despeckling results compared to SAR-DRN, a state-of-the-art CNN-based algorithm.
\end{quote}
\end{abstract}

\section{Introduction}
Multiplicative noise, also known as speckle noise, typically occurs in active imaging system, for example, laser images, microscope images, and SAR (Synthetic Aperture Radar) images. While many general image denoising algorithms have focused on the additive noise setting, despeckling the multiplicative noise is also becoming important as the active sensors, e.g., SAR, are gradually becoming a significant source of remote sensing data in the field of geographic mapping, resource surveying, and military reconnaissance, etc. 

During the last few decades, various different approaches have been proposed for the multiplicative noise despeckling; \emph{e.g.}, Bayesian methods \cite{lee1980digital,LEE1981380}, non-local filtering \cite{buades2005non,dabov2007image,rudin1992nonlinear}, total variation regularization \cite{huang2009new}, compressed sensing based \cite{HaoFenXu12}, and variational approaches \cite{AubAuj08}, etc. For more extensive survey on the topic, we refer the readers to \cite{argenti2013tutorial}. 

In addition to the above mentioned classical approaches, the deep learning-based multiplicative noise despeckling methods have been recently considered, as in the additive noise denoising case, \emph{e.g.}, \cite{zhang2017beyond}. Namely, by casting the despeckling as a supervised learning problem, the CNN models are trained to learn a mapping from the noisy patch to the clean patch. For example, \cite{wang2017sar} proposed ID-CNN that uses exactly the same architecture as DnCNN on the Gamma-distributed multiplicative noise and introduced a total variation (TV) regularization; \cite{chierchia2017sar} transformed both clean and noisy images to the log domain then carried out the same process as in the additive noise case; \cite{zhang2018learning} proposed SAR-DRN, which added dilated convolution and skip connections to the ID-DCN architecture. As in the additive noise denoising case, the CNN-based models have achieved superior despeckling performances than the classical approaches. 

Despite the impressive performance, above methods have one critical limitation, however. That is, once the supervised training of the network is done, the network parameters get frozen and no adaptation of the despeckling model to a given noisy image can be done. To overcome such limitation, \cite{cha2018neural} have recently proposed an adaptive method, dubbed as N-AIDE, for the \emph{additive noise} case that can carry out both the supervised training (on an offline dataset) and adaptive fine-tuning (on a given noisy data) of the network parameters. Later, N-AIDE was extended in \cite{cha2018fully} to implement a fully convolutional architecture and unknown noise variance estimation scheme. The crux of N-AIDE
is to design the neural network to learn pixelwise affine mappings with a specific conditional independence constraint and learn the network parameters by setting an unbiased estimate of the true mean-squared error (MSE) of the mappings as an optimizing objective. 

In this paper, we show the framework of N-AIDE can be successfully extended to the multiplicative noise case as well and attain the state-of-the-art performance. Our contribution is threefold. First, we derive a new unbiased estimate of MSE for the multiplicative noise case while remaining in learning the pixelwise affine mappings with neural networks, as in N-AIDE. Our estimate of MSE can be interpreted to give an intuitive explanation of the adaptivity of our algorithm. Second, we devise a novel double-sided masked CNN architecture, dubbed as DoPAMINE (\underline{Do}uble-sided masked CNN for \underline{P}ixelwise \underline{A}daptive \underline{M}ult\underline{I}plicative \underline{N}oise d\underline{E}speckling), which maintains the conditional independence property of each pixel in any feature map given the surrounding input pixels. Such property is indispensable for applying the framework of N-AIDE, and obtaining such property with CNN architecture is not trivial as we elaborate below. Third, we propose a \emph{scale add} layer within the DoPAMINE architecture and show such simple layer can accelerate the training of our model that contains many add operations among the intermediate feature maps. Compared to the popular He initialization \cite{he2015delving} or Batch Normalization \cite{ioffe2015batch}, we show our Scale Add layer leads to a much stable and fast training of the base supervised model. Combining the three contributions, we show that our DoPAMINE significantly outperforms SAR-DRN \cite{zhang2018learning}, the current CNN-based state-of-the-art despeckling model, on a benchmark dataset. Furthermore, from real SAR image despeckling results, we show DoPAMINE has a knob to control the emphasis on either homogeneous regions or sharp edges and present that it can capture finer details of the images compared to SAR-DRN.

\section{Notations and Preliminaries}
\subsection{Multiplicative noise model}
We denote the vectorized original clean data and noise as $\mathbf{x}\in\mathbb{R}^d$ and $\mathbf{N}\in\mathbb{R}^d$, respectively. The noisy observation is then denoted as $\Z \in \mathbb{R}^d$, and the $i$-th element of $\Z$ is defined as

\begin{align}
Z_i = x_iN_i,\ \ \ i=1,2,\ldots,d\label{eq:mult_noise}
\end{align}
in which we assume $N_i$'s are independent with $\mathbb{E}(N_i)=1$ and $\text{Var}(N_i)=\sigma^2$ for all $i$. We do not assume $N_i$ to be identically distributed. We use uppercase letters to denote random variables or random vectors and do not make any stochastic assumptions on the original data $\mathbf{x}$. In the multiplicative noise despeckling literature, $N_i$'s are typically assumed to follow the Gamma distribution, but in this paper, we \emph{do not} require the noise to be Gamma-distributed.

The reconstructed data after despeckling is denoted as $\xhat(\Z)=\{\hat{x}_i(\Z)\}_{i=1}^d$. Note the notation emphasizes the dependency on the entire $\Z$ for obtaining the reconstruction of the $i$-th element. The goodness of despeckling is typically measured with the mean-squared error (MSE) as in the additive noise case. 

\subsection{N-AIDE and the unbiased  estimator}
\cite{cha2018neural} considered the additive noise denoising case and suggested to use the pixelwise affine mappings to obtain the reconstructions:
\begin{align}
\hat{x}_i(\textbf{Z}) = a(\Zi)Z_i+b(\Zi),\ \ \ i=1,2,\ldots,d\label{eq:affine_mappings}
\end{align}
in which $\Zi$ denote the entire noisy data except for the $i$-th element. Namely, in (\ref{eq:affine_mappings}), the slope and bias constants for the $i$-th location become conditionally independent of $Z_i$ given $\Zi$ (due to the independence of noise and no stochastic assumption on $\x$). Now, to simplify the notation, we use $a_i$ and $b_i$ to denote $a(\Zi)$ and $b(\Zi)$, respectively. Furthermore, we denote $\mathbf{a}\in\mathbb{R}^d$ and $\mathbf{b}\in\mathbb{R}^d$ as vectorized slope and bias constants, hence, we can also express $\xhat(\Z)=\mathbf{a}\odot \Z+\mathbf{b}$, in which $\odot$ stands for the element-wise multiplication.

Using the conditional independence property, \cite{cha2018neural} devised an unbiased estimate of MSE for (\ref{eq:affine_mappings}) in the additive noise case, i.e., 
\begin{align}
    \mathbf{L}_{\text{add}}(\Z,(a_i,b_i);\sigma^2) = (Z_i-\hat{x}_i(\Z))^2+\sigma^2(2a_i-1),\label{eq:additive esimate}
\end{align}
which is shown to satisfy 
\begin{align}
&\mathbb{E}_{Z_i}\big[(x_i-\hat{x}_i(\Z))^2|\Zi\big] \nonumber \\ 
&=\mathbb{E}_{Z_i}\big[\mathbf{L}_{\text{add}}(\Z,(a_i,b_i);\sigma^2) |\Zi\big].\label{eq:additive unbiased}
\end{align}
With above property, N-AIDE used fully-connected neural network to output $a_i$ and $b_i$ based on $\mathbf{C}_{k\times k}^{-i}$, the $k\times k$ two dimensional context patch with a hole at location $i$. The network parameters of N-AIDE were trained first by supervised training using the regular MSE and a separate supervised training set, then by adaptive fine-tuning using (\ref{eq:additive esimate}) and a given noisy image.

\begin{figure*}[ht]
    \centering
    \begin{subfigure}[b]{0.95\textwidth}
        \includegraphics[width=\textwidth]{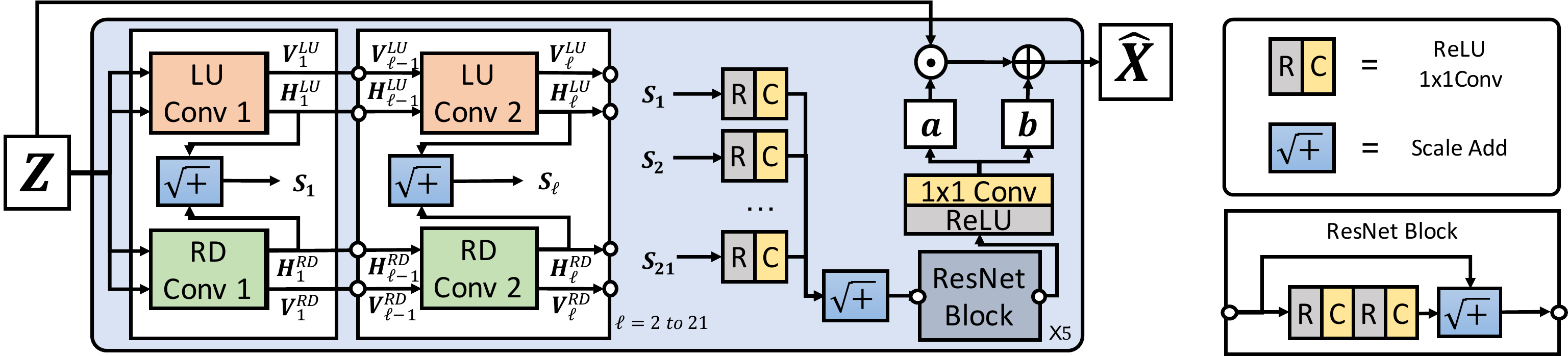}
        \caption{An overall architecture of DoPAMINE.}
    \end{subfigure}
    
    \begin{subfigure}[b]{0.28\textwidth}
        \includegraphics[width=\textwidth]{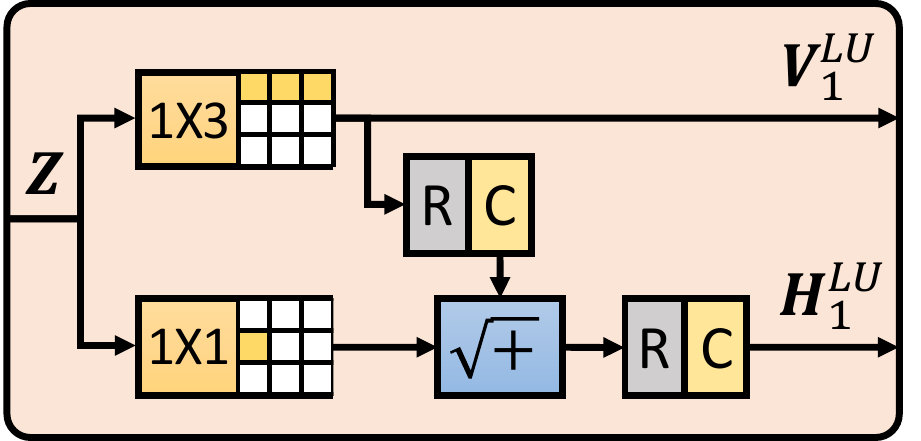}
        \caption{A LU Conv 1 block.}
    \end{subfigure}
        \begin{subfigure}[b]{0.36\textwidth}
        \includegraphics[width=\textwidth]{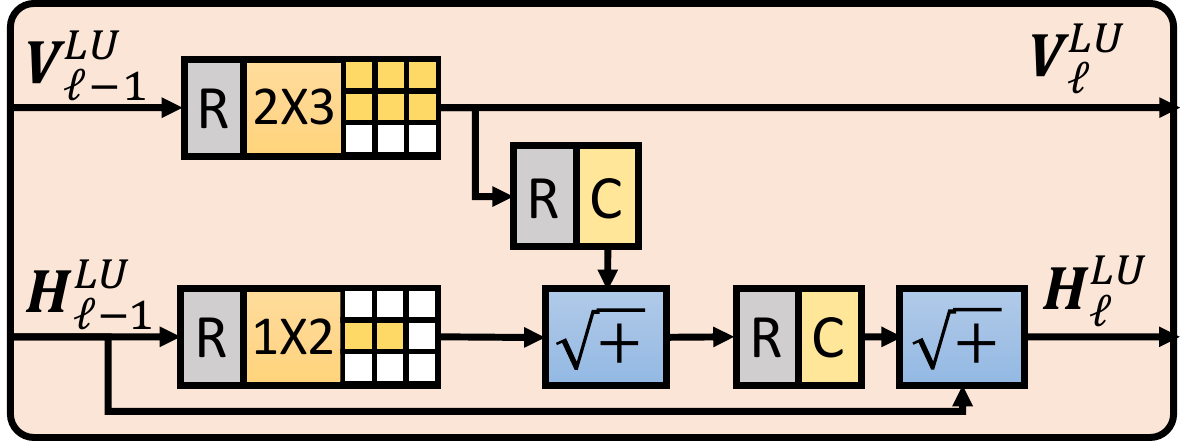}
        \caption{A LU Conv 2 block. It repeated 20 times.}
    \end{subfigure}
        \begin{subfigure}[b]{0.30\textwidth}
        \includegraphics[width=\textwidth]{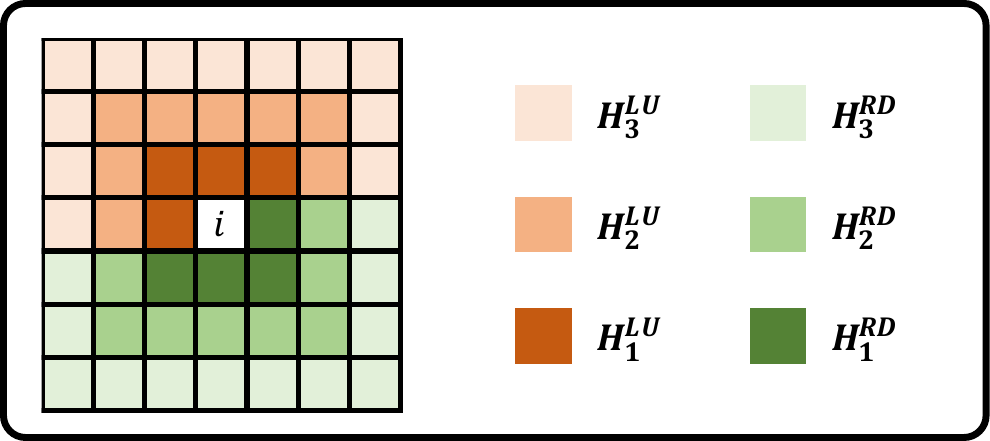}
        \caption{Receptive field of horizontal stack.}
    \end{subfigure}
    \caption{Full representation for DoPAMINE model. Best viewed in color.}
    \label{fig:DoPAMINE}
\end{figure*}

\section{Proposed Method}
\subsection{Unbiased estimator of MSE for multiplicative noise}

Following N-AIDE, we obtain the following lemma for the multiplicative noise case. 

\begin{lemma}
For the affine mapping $\hat{x}_i(\Z)$ defined in (\ref{eq:affine_mappings}) and the multiplicative noise case in (\ref{eq:mult_noise}),
\begin{align}
    \mathbf{L}_{\text{mult}}(\Z,(a_i,b_i);\sigma^2) = (Z_i-\hat{x}_i(\Z))^2+\frac{Z_i^2\sigma^2}{1+\sigma^2}(2a_i-1)\label{eq:multiplicative esimate}
\end{align}
becomes an unbiased estimate of MSE, $(x_i-\hat{x}_i(\Z))^2$. 
\end{lemma}
\emph{Proof:} The proof follows from the conditional independence of $(a_i,b_i)$ and $Z_i$ given $\Z^{-i}$. \qed

Given (\ref{eq:multiplicative esimate}), the training process of a neural network follows that of N-AIDE. Namely, assuming the noise distribution is known, as is the case in other despeckling algorithms, we first generate a supervised training dataset that contains both clean and multiplicative noise-corrupted images. Then, we define a neural network that outputs $a_i$ and $b_i$ for the input $\mathbf{C}_{k\times k}^{-i}$ and train its parameters with the supervised training set and the MSE loss function. The specific architecture of our neural network is given in the following section. After the supervised training, given a noisy image $\Z\in\mathbb{R}^d$ subject to despeckling, we define the  adaptive loss function as
\begin{align}
    \mathcal{L}_{\text{FT}}(\Z,\sigma)\triangleq \frac{1}{d}\sum_{i=1}^d\mathbf{L}_{\text{mult}}(\Z,(a_i,b_i);\sigma^2)\label{eq:fine_tune_loss}
\end{align}
to fine-tune the network parameters by minimizing it. As in N-AIDE, such fine-tuning achieves strong adaptivity of our method, which we highlight in the experimental section. Furthermore, in order to strengthen the adaptivity, we also define the data augmented fine-tuning (AFT) loss as in \cite[Section 3.2]{cha2018fully},
\begin{align}
    \mathcal{L}_{\text{AFT}}(\Z,\sigma) = \frac{1}{8}\sum_{\Z^{(j)}\in\mathcal{A}(\Z)}\mathcal{L}_{\text{FT}}(\Z^{(j)},\sigma),\label{eq:aft_loss}
\end{align}
in which $\mathcal{A}(\Z)$ stands for the augmented dataset that consists of flipped and $(0^{\circ},90^{\circ},180^{\circ},270^{\circ})$-degrees rotated versions of $\Z$. Once the fine-tuning process is done, the noisy image $\Z$ is then despeckled by applying the learned affine mapping $(a_i,b_i)$ for each location.

\noindent\emph{Remarks:}
Although we are using the same pixelwise affine function form (\ref{eq:affine_mappings}) and the framework of N-AIDE, comparing (\ref{eq:multiplicative esimate}) with (\ref{eq:additive esimate}) clearly shows how the affine mappings are learned differently during fine-tuning for the multiplicative noise case compared to the additive noise case. Namely, in both estimators, we can interpret the second terms in (\ref{eq:multiplicative esimate}) and (\ref{eq:additive esimate}) as some sort of regularizations on the slope parameter $a_i$. That is, while the first squared error term will be minimized with $a_i=1$ and $b_i=0$, due to the existence of the second term, $a_i$ should shrink and the residual, $(1-a_i)Z_i$, would be corrected with $b_i$. Now, we see that the ``effective'' regularization parameter in (\ref{eq:additive esimate}) is fixed to a constant $\sigma^2$ for all $i$, whereas in (\ref{eq:multiplicative esimate}), it depends on $Z_i^2$. Hence, for the pixels with large $Z_i^2$ values, our model should output smaller slope parameter $a_i$ to minimize $\mathbf{L}_{\text{mult}}(\Z,(a_i,b_i);\sigma^2)$ than the pixels with small $Z_i^2$ values, and vice versa. Note this tendency makes perfect sense for the multiplicative noise despeckling. Namely, since the noise gets multiplied to the original $x_i$ as in (\ref{eq:mult_noise}), the noise level may amplify significantly to result in very large $Z_i$ when $x_i$ has high intensity and $N_i>1$. In such case, it is reasonable to suppress the noise effect by shrinking $a_i$ and make the corrections with the bias term $b_i$ to accurately estimate $x_i$. In our experimental results below, we show how the learned $a_i$ and $b_i$ vary depending on the intensity of $Z_i$'s and how our adaptive fine-tuning makes corrections to achieve better despecklings.

\begin{table*}[ht]
    \centering
    \caption{PSNR and SSIM on UCML dataset.}
    \begin{tabular}{|c|*{4}{|c}|}
    \hline
\multirow{2}{*}{Model} & L=1.0 & L=2.0 & L=4.0 & L=8.0 \\ \cline{2-5}
                 & PSNR(dB) / SSIM & PSNR(dB) / SSIM & PSNR(dB) / SSIM & PSNR(dB) / SSIM \\ \hline
Noise image         & 5.763 / 0.073 & 8.773 / 0.118 & 11.785 / 0.178 & 14.795 / 0.253\\ \hline
SAR-DRN$_{S}$\cite{zhang2018learning}      & 22.86 / 0.738 & 24.189 / 0.780 & 25.664 / 0.823 & 27.191 / 0.861\\ \hline
SAR-DRN$_{B}$     & 22.746 / 0.732 & 24.151 / 0.778 & 25.583 / 0.821 & 27.057 / 0.857\\ \hline
DoPAMINE$_{S}$     & 22.957 / 0.743 & 24.329 / 0.786 & 25.767 / 0.826 & 27.268 / 0.863\\ \hline
DoPAMINE$_{S-AFT}$ & \textbf{23.308} / \textbf{0.756} & \textbf{24.701} / \textbf{0.798} & \textbf{26.123}  / \textbf{0.837} & \textbf{27.623} / \textbf{0.871}\\ \hline
DoPAMINE$_{B}$     & 22.909 / 0.739 & 24.291 / 0.782 & 25.694 / 0.822 & 27.159 / 0.859\\ \hline
DoPAMINE$_{B-AFT}$ & 23.293 / 0.755 & 24.678 / 0.797 & 26.103 / 0.836 & 27.593 / 0.870\\ \hline
    \end{tabular}
    \label{tab:PSNR}
\end{table*}

% \begin{table*}[ht]
%     \centering
%     \caption{PSNR and SSIM on UCML dataset.}
%     \begin{tabular}{|c|*{8}{|c}|}
%     \hline
% \multirow{2}{*}{Model} & \multicolumn{2}{c|}{L=1.0} & \multicolumn{2}{c|}{L=2.0} & \multicolumn{2}{c|}{L=4.0} & \multicolumn{2}{c|}{L=8.0} \\ \cline{2-9}
%                  & PSNR(dB) & SSIM & PSNR(dB) & SSIM & PSNR(dB) & SSIM & PSNR(dB) & SSIM \\ \hline
% Noise image         & 5.763 & 0.073 & 8.773 & 0.118 & 11.785 & 0.178 & 14.795 & 0.253\\ \hline
% SAR-DRN$_{S}$\cite{zhang2018learning}      & 22.86 & 0.738 & 24.189 & 0.78 & 25.664 & 0.823 & 27.191 & 0.861\\ \hline
% SAR-DRN$_{B}$     & 22.746 & 0.732 & 24.151 & 0.778 & 25.583 & 0.821 & 27.057 & 0.857\\ \hline
% DoPAMINE$_{S}$     & 22.957 & 0.743 & 24.329 & 0.786 & 25.767 & 0.826 & 27.268 & 0.863\\ \hline
% DoPAMINE$_{S-AFT}$ & \textbf{23.308} & \textbf{0.756} & \textbf{24.701} & \textbf{0.798} & \textbf{26.123}  & \textbf{0.837} & \textbf{27.623} & \textbf{0.871}\\ \hline
% DoPAMINE$_{B}$     & 22.909 & 0.739 & 24.291 & 0.782 & 25.694 & 0.822 & 27.159 & 0.859\\ \hline
% DoPAMINE$_{B-AFT}$ & 23.293 & 0.755 & 24.678 & 0.797 & 26.103 & 0.836 & 27.593 & 0.87\\ \hline
%     \end{tabular}
%     \label{tab:PSNR}
% \end{table*}

%\input{4_Architecture.tex}
\subsection{DoPAMINE architecture}

As we mentioned in (\ref{eq:affine_mappings}), the main critical constraint that ensures the unbiasedness of (\ref{eq:multiplicative esimate}) is that for every pixel $i$, the $a_i$ and $b_i$ should be conditionally independent of $Z_i$ given $\Zi$. In N-AIDE \cite{cha2018neural}, a fully connected neural network was used, but due to the architectural simplicity, it could not outperform other deep learning baselines (which do not possess the adaptivity) that use convolutional architectures. Recently, in \cite{cha2018fully}, a nontrivial extension that uses fully \emph{convolutional} architecture was proposed and the resulting algorithm was shown to outperform several strong CNN-based state-of-the-arts for the additive noise case. The main challenge of using the fully convolutional architecture is that when a vanilla architecture, e.g., FCN \cite{long2015fully}, is used, the $i$-th pixel in a feature map may depend on $Z_i$, which would result in breaking the critical constraint mentioned above. 

Here, we independently propose another fully convolutional architecture, DoPAMINE, that is summarized in Figure \ref{fig:DoPAMINE}. Two key ingredients of our architecture are the LU convolution, which is inspired by the PixelCNN \cite{van2016conditional}, and the scale add layer, which is simple but very effective in expediting the training of CNN models with many addition operations among the feature maps, like ResNet \cite{he2016deep}.

\subsubsection{LU Convolution}
The PixelCNN in \cite{van2016conditional} was developed as a generative model that can sequentially and conditionally generate images. The main gist in their model was to devise a masked convolution architecture that can generate each pixel conditioned on the pixels that has ``causal'' relationship with the pixel. In contrast, in our despeckling problem, we are not directly generating the reconstruction for pixel $i$, but are estimating $a_i$ and $b_i$ for the affine mapping, and the estimation is based on the ``double-sided, non-causal'' context of the pixel.

Hence, as shown in Figure \ref{fig:DoPAMINE}(a), we adopt the masked convolution architecture of PixelCNN twice in each layer $\ell$ to cover both the causal part (LU convolution) and the anti-causal part (RD convolution) for each pixel $i$. We then ``scale add'' (denoted as $\sqrt{+}$) the two feature maps from LU and RD convolutions to generate the resulting feature map $S_{\ell}$. By this construction, 
one can see that any pixel $i$ on $S_{\ell}$ does not depend on $Z_i$ and is computed only based on the double-sided context (with corresponding receptive field) of $Z_i$. 
Once we obtain $S_{\ell}$, we then repeatedly stack the LU and RD convolution layers up to $\ell=21$.  
Now, since the RD convolution is just a $180^{\circ}$ rotated version of LU convolution, we just elaborate on the LU convolution more in details below.   

As shown in Figure \ref{fig:DoPAMINE}(b)(c), the structure of LU convolution for layer 1 and layers $\ell=2\sim21$ are different. In LU Conv 1, the masked $1\times 3$ and $1\times 1$ filters are used to generate the horizontal stack feature map, $H_{1}^{LU}$, of which receptive field of the $i$-th pixel is given in Figure \ref{fig:DoPAMINE}(d). Note the ``scale add'' was also used to combine the feature maps from the two filters. The LU Conv 2 then uses differently masked filters such that the ``causal'' receptive fields can grow as layer increases as depicted in Figure \ref{fig:DoPAMINE}. The RD Conv Layers operates in the same way, hence, the final receptive field of a pixel in $S_{21}$ is $43\times43$.

Finally, we ``scale add'' all the feature maps from all the layers before passing them through the ResNet block depicted in Figure \ref{fig:DoPAMINE}. While PixelCNN only uses the last horizontal stack, we chose to directly use all the feature maps to use the low level features in our despeckling problem. 

\subsubsection{Scale add layer}
He initialization \cite{he2015delving} and Batch Normalization (BN) \cite{ioffe2015batch} are widely used practice to accelerate training speed by good initialization and reducing covariate shift.
However, both methods also have some drawbacks. Namely, He initialization turns out to be disharmonious with addition layer, since it may significantly increase the variance of the resulting feature maps. This could be problematic in modern CNN architectures that have many addition layers among feature maps, \emph{e.g.}, ResNet \cite{he2016deep} or DenseNet \cite{huang2017densely}, as it may hinder fast training of the network. Moreover, batch normalization requires additional memory, and it is known to be not proper for non-i.i.d. or small sized mini-batches \cite{ioffe2017batch}. In addition, in standard models for pixel-wise reconstruction problems, \textit{e.g.}, FCN, PixelCNN and WaveNet\cite{van2016wavenet}, the batch normalization does not show as significant improvements as in the classification problems. 

To address above drawbacks, we propose a simple \emph{scale add layer}, which can replace addition layers while preserving the variances of feature maps without introducing any additional parameters, memory, and batch statistics, as in batch normalization.
We denote $N$ number of feature maps as $\{\mathbf{Y}_i\}_{i=1}^N$. Then, the scale add layer, $SA$, is defined as:
\begin{align}
SA(\mathbf{Y}_1, \mathbf{Y}_2, ..., \mathbf{Y}_N) = \frac{1}{\sqrt{N}}\sum_{i=1}^N \mathbf{Y}_i.
\end{align}
With this simple scaling, reminiscent of the motivations of He and Xavier initialization, $SA$ equalizes the input and output feature map variances, while a simple addition layer will increase variance $N$ times.
Note if there are $L$ addition layers, then the order of variance for the final layer's feature map becomes $N^L$ \cite{he2015delving}.
In contrast, by assuming $Var[\mathbf{Y_i}]\approx C$  for all $i$ and all the feature maps are uncorrelated, the output variance of $SA$ can be calculated as:
\begin{align*}
&\text{Var}[SA(\mathbf{Y}_1, \mathbf{Y}_2, ..., \mathbf{Y}_N)] \\
& = \text{Var}[\frac{1}{\sqrt{N}}(\sum_{i=1}^N \mathbf{Y}_i)] \approx \frac{1}{N}\sum_{i=1}^N \text{Var}[\mathbf{Y_i}] = C.
\end{align*}
Thus, when combined with He initialization, which is known to control the variances of feature maps from ReLU activated convolution layers, one can see that the simple $SA$ can maintain the variance of the feature maps resulting from any number of additions. In the experimental section, we convincingly show this point.

\begin{figure*}[ht]
    \centering
    \makebox[20pt]{\raisebox{40pt}{\rotatebox[origin=c]{90}{Flevoland}}}%
    \begin{subfigure}[b]{0.18\textwidth}
        \includegraphics[width=\textwidth]{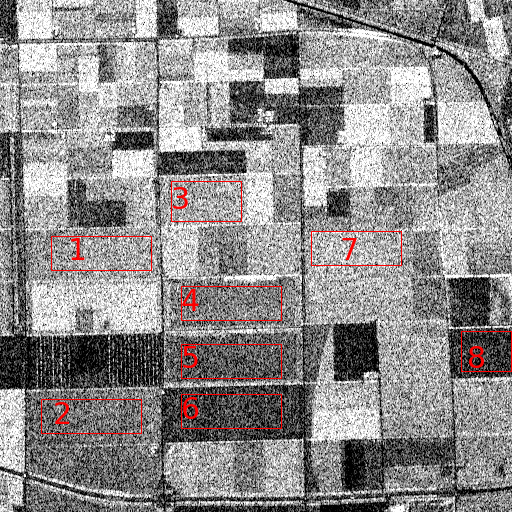}
        \caption{Noise image}
    \end{subfigure}
    \begin{subfigure}[b]{0.18\textwidth}
        \includegraphics[width=\textwidth]{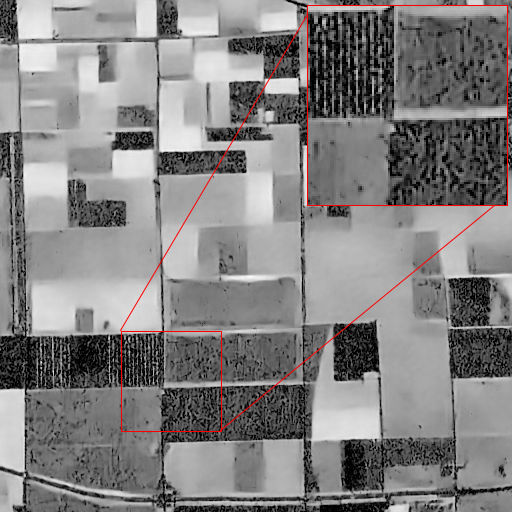}
        \caption{SAR-DRN$_S$}
    \end{subfigure}
    \begin{subfigure}[b]{0.18\textwidth}
        \includegraphics[width=\textwidth]{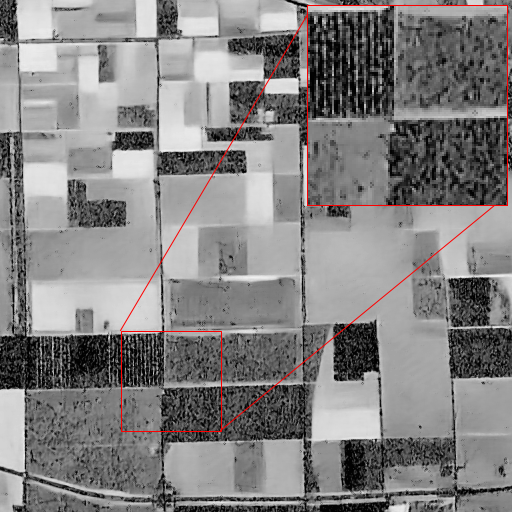}
        \caption{DoPAMINE$_S$}
    \end{subfigure}
    \begin{subfigure}[b]{0.18\textwidth}
        \includegraphics[width=\textwidth]{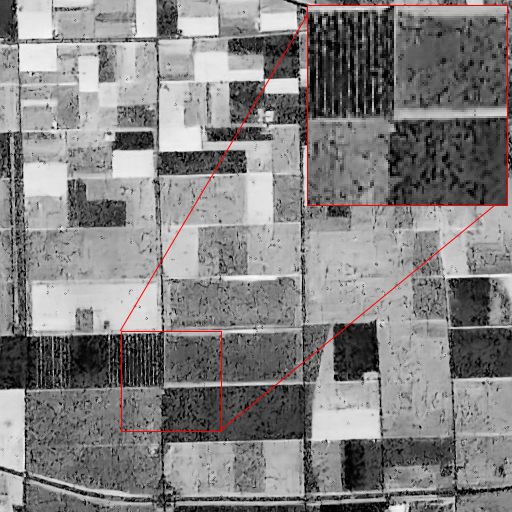}
        \caption{DoPAMINE$_B$}
    \end{subfigure}
    \begin{subfigure}[b]{0.18\textwidth}
        \includegraphics[width=\textwidth]{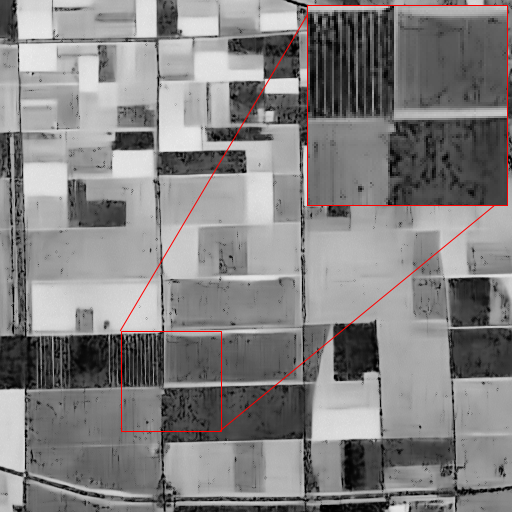}
        \caption{DoPAMINE$_{B-AFT}$}
    \end{subfigure}
    
    \caption{Noise and despeckled Flevoland images. We trained supervised model with $L=4$, which is same as the number of look of image.}
    \label{fig:realSar}
\end{figure*}

\begin{table*}[ht]
    \centering
    \caption{ENL of noise image and despeckled images.}
\begin{tabular}{|c|*{8}{|c}||c|}
\hline
Model              & ENL1           & ENL2           & ENL3           & ENL4            & ENL5           & ENL6           & ENL7          & ENL8          & Avg            \\ \hline
Noise image        & 5.3            & 11.0           & 11.9            & 7.4            & 3.6            & 2.2           & 13.4           & 1.7           & 7.1            \\ \hline
DoPAMINE$_{S}$     & 34.8           & \textbf{573.7} & \textbf{1351.6} & 260.0          & 14.6           & 5.1           & \textbf{885.2} & 3.7           & 391.1          \\ \hline
DoPAMINE$_B$       & 23.6           & 120.7          & 131.3           & 44.0           & 23.2           & 12.0          & 226.5          & 11.5          & 74.1           \\ \hline
SAR-DRN$_{S}$      & 62.4           & 518.5          & 1284.3          & \textbf{476.2} & 15.6           & 4.5           & 689.1          & 3.2           & 381.7          \\ \hline
SAR-DRN$_B$        & 19.0           & 83.4           & 73.6            & 37.1           & 13.3           & 5.8           & 130.7          & 5.9           & 46.1           \\ \hline
DoPAMINE$_{B-AFT}$ & \textbf{142.5} & 531.0          & 1258.4          & 412.5          & \textbf{141.8} & \textbf{36.0} & 793.2          & \textbf{32.0} & \textbf{418.4} \\ \hline
\end{tabular}
    \label{tab:ENL}
\end{table*}

\section{Experiment}
\subsection{Benchmark dataset}
\subsubsection{Training detail}
We use UC Merced Land use dataset (UCML) \cite{yang2010bag} to train and test our model. The dataset contains 21 classes, and each class has 100 images with the resolution of $256\times256$. We chose the Airplane, River, and Building classes as our test set. For the supervised training set, we randomly sampled 400 images from the remaining 18 classes. Then, we cropped the images to 40$\times$40 patch size with stride 10, so that the total number of image patches in the supervised training set is 193,600. The batch size of supervised learning ($S$) and data augmented fine tuning ($AFT$) were 64 and 1, respectively. In contrast to $S$, we put full 256$\times$256 resolution of image during $AFT$.
The number of filters in each convolution layer of DoPAMINE was 64, except for last convolution layer that has two $1\times 1$ filters. Learning rate for $S$ was initially set to $10^{-3}$ and got halved for every 10 epochs. Learning rate for $AFT$ was $1.2\times10^{-5}$. The number of epochs for $S$ and $AFT$ were 30 and 10, respectively. 

For generating the multiplicative noise, we used the Gamma distribution parameterized by $L$, which has the following density function
\begin{align*}
    p_{N_i}(n_i) = \frac{L^Ln_i^{L-1}e^{-n_i L}}{\Gamma(L)}.
\end{align*}
Note $E[N_i]$ and $Var[N_i]$ are $1$ and $1/L$, respectively. We tested on the benchmark dataset with four noise levels, $L=1,2,4,$ and $8$.
To avoid overfitting, we did the noise-augmented training during supervised learning, namely, randomly generated $N_i$ for every epoch and constructed new realization of noisy patches. Adam optimizer\cite{kingma2014adam} is used for both $S$ and $AFT$. Our model is implemented by Keras with Tensorflow backend, and trained on NVIDIA GTX 1080Ti with CUDA 9.2.

\subsubsection{Blind model}
In addition to the ordinary supervised model for each noise level, we also train a Blind model, of which weights are trained with various noise variances. Such blindly trained supervised model was first proposed in \cite{zhang2017beyond}. For training the blind model, we split cropped training images such that each group has 121 cropped images. Then, we generated different Gamma distributions for each group with $L$ sampled from  $U[0.5,12]$ to generate training data with multiple noise levels. We then shuffled all the images before training and carried out the ordinary mini-batch training with Adam. 

Such blindly trained model is well-known to be robust to multiple noise levels. In our experimental results below, we show that when combined with the adaptive fine-tuning step, the blind model can be made as strong as the supervised model matched to a specific noise level.

\subsubsection{Metrics}
We use PSNR and SSIM, commonly used metrics, to evaluate the despeckling performances. The PSNR between $\textbf{x}\in[0,1]^d$ and $\mathbf{\hat{x}}$ is defined by:
\begin{align*}
    PSNR(\mathbf{x}, \mathbf{\hat{x}}) = 10 \log_{10} \frac{1}{\sum_{i=1}^d (x_i-\hat{x}_i)^2/d}
\end{align*}
Since the denominator is MSE between $\textbf{x}$ and $\mathbf{\hat{x}}$, the higher PSNR, the better.

Structural similarity, SSIM, is another metric defined as
\begin{align*}
    SSIM(\mathbf{x}, \mathbf{\hat{x}}) = \frac{(2\mu_x\mu_{\hat{x}}+c_1)(2\sigma_{x{\hat{x}}}+c_2)}{(\mu_x^2+\mu_{\hat{x}}^2+c_1)(\sigma_x^2+\sigma_{\hat{x}}^2+c_2)}
\end{align*}
and is designed to cover the weakness of PSNR, which is known to be sensitive to shift or brightness of the images

\begin{figure*}[ht]
    \centering
    \begin{subfigure}[t]{0.305\textwidth}
        \includegraphics[width=\textwidth]{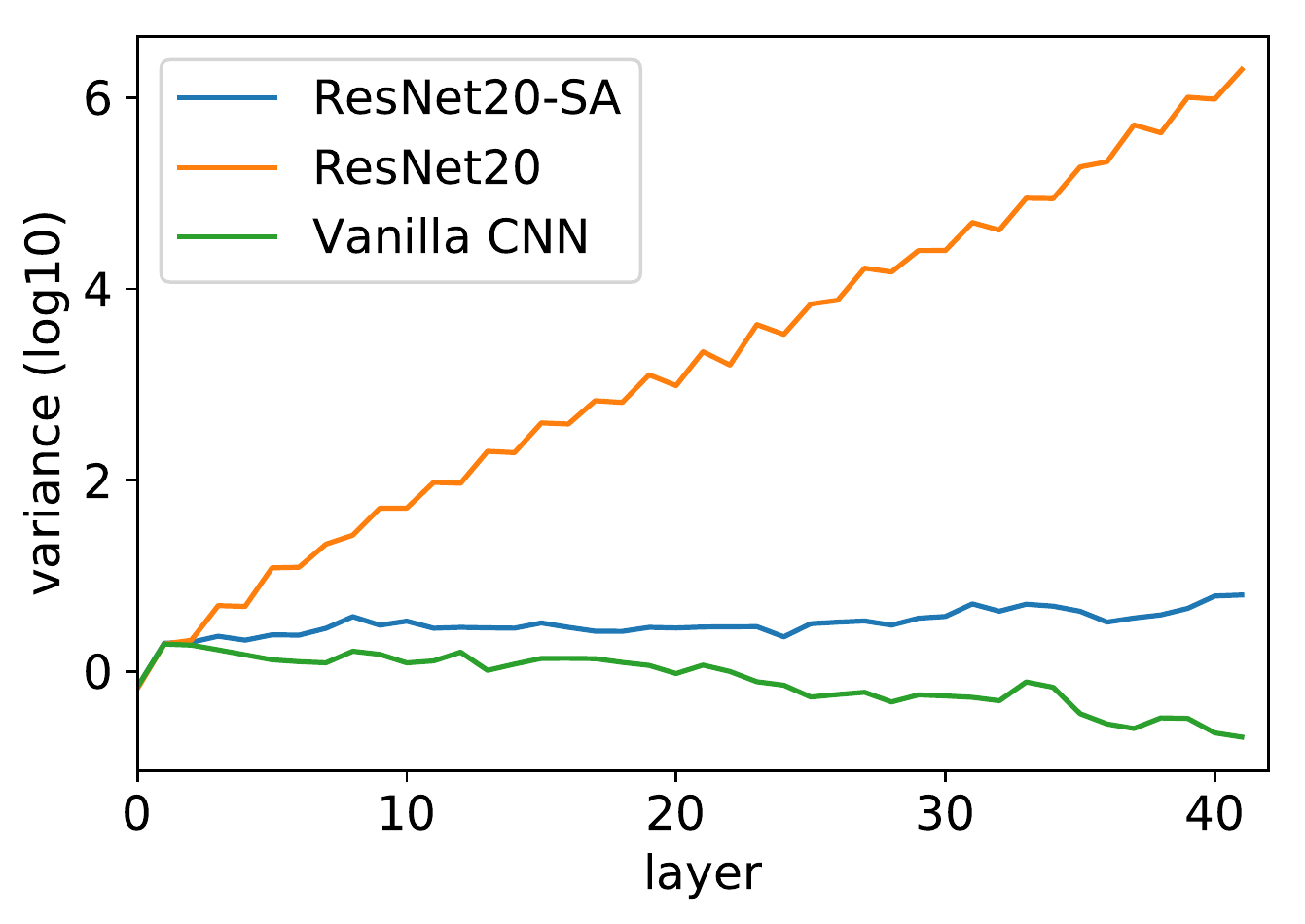}
        \caption{Initial variance of each conv layer in ResNet20 for Gaussian input.}
    \end{subfigure}
    \begin{subfigure}[t]{0.33\textwidth}
        \includegraphics[width=\textwidth]{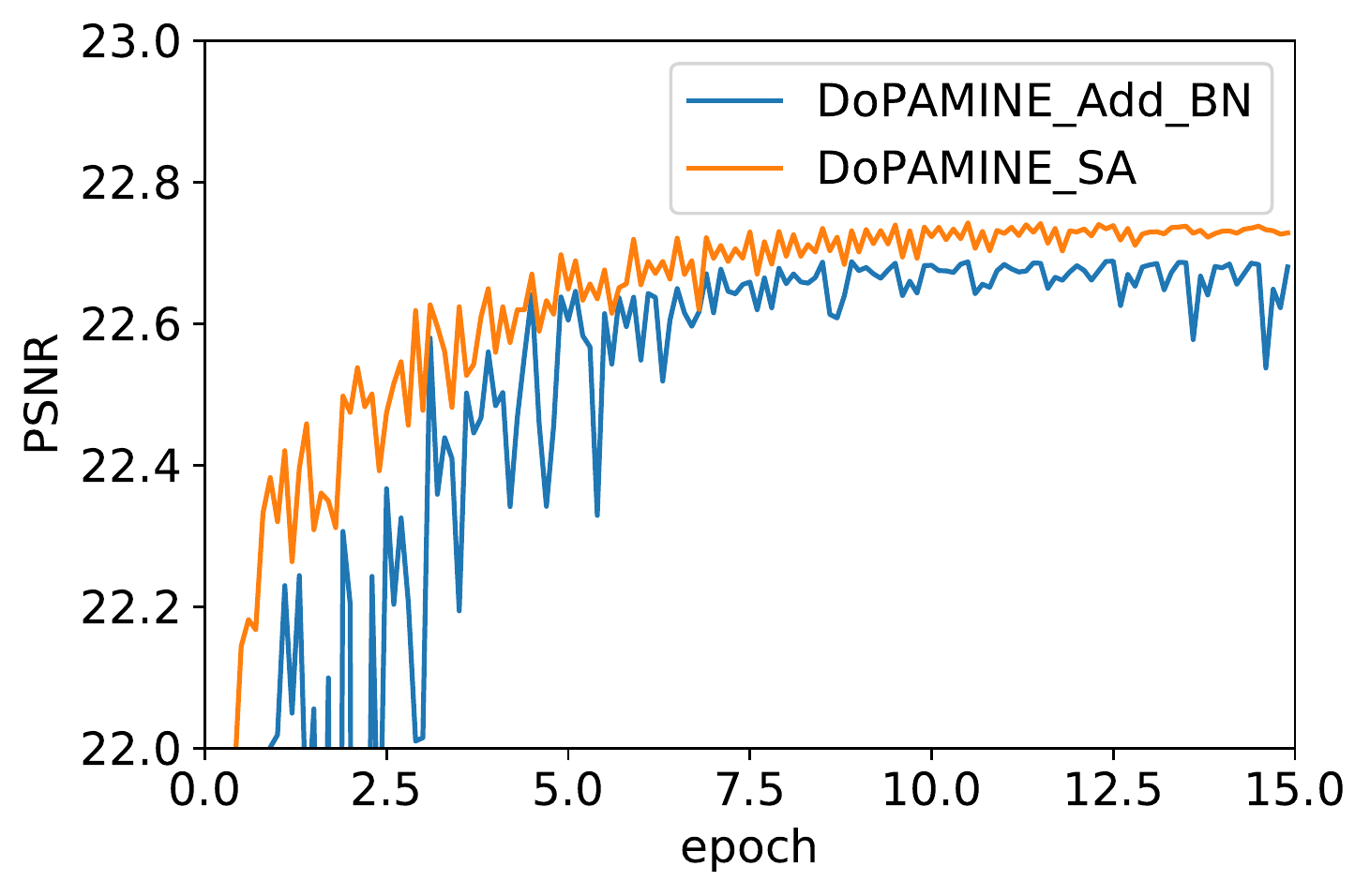}
        \caption{The comparison of DoPAMINE\_SA and DoPAMINE\_Add\_BN on UCML dataset.}
    \end{subfigure}
    \centering
    \begin{subfigure}[t]{0.32\textwidth}
        \includegraphics[width=\textwidth]{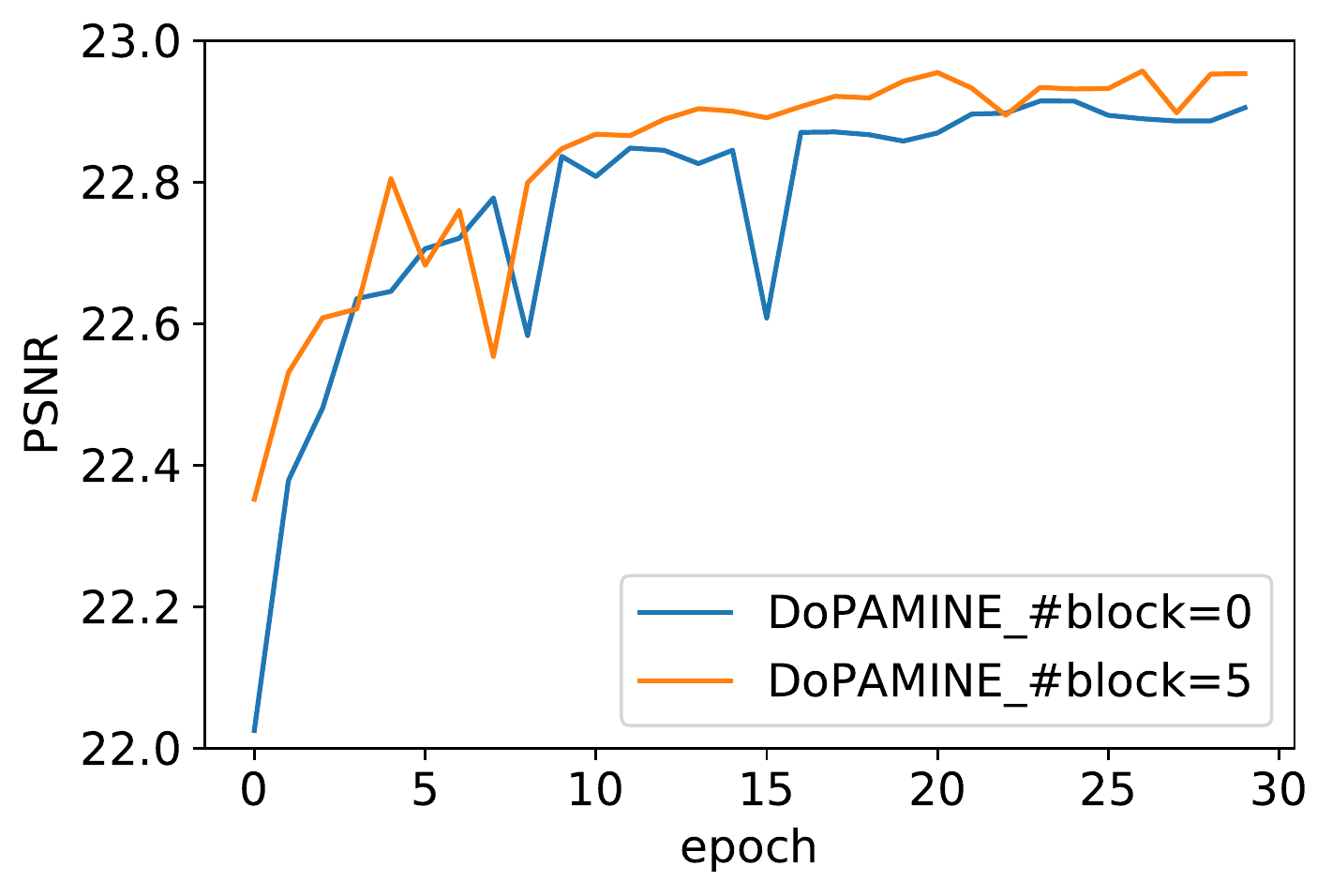}
        \caption{Existence of ResNet blocks.}
    \end{subfigure}
    
    \begin{subfigure}[t]{0.32\textwidth}
        \includegraphics[width=\textwidth]{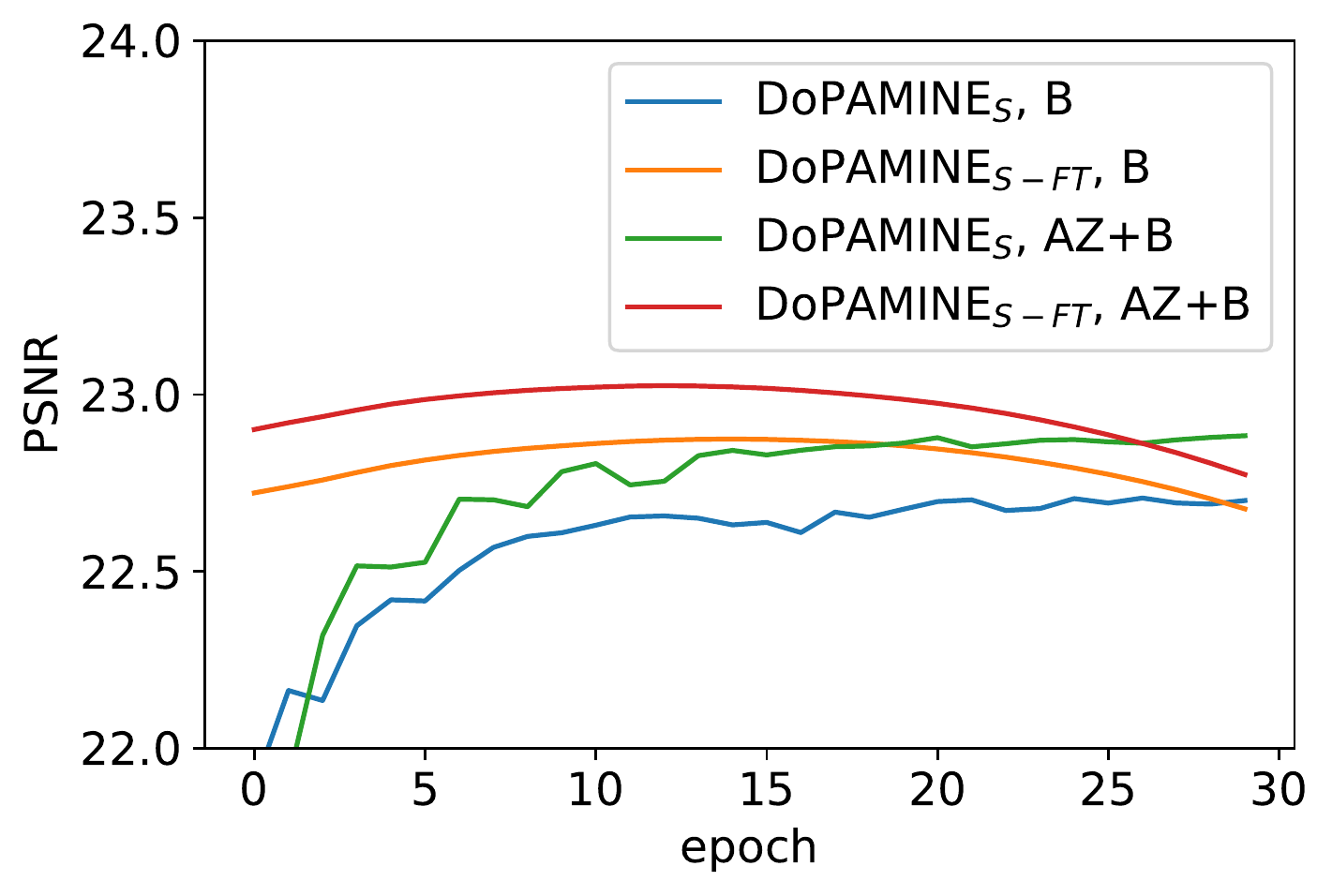}
        \caption{Affine mapping}
    \end{subfigure}
    \begin{subfigure}[t]{0.32\textwidth}
        \includegraphics[width=\textwidth]{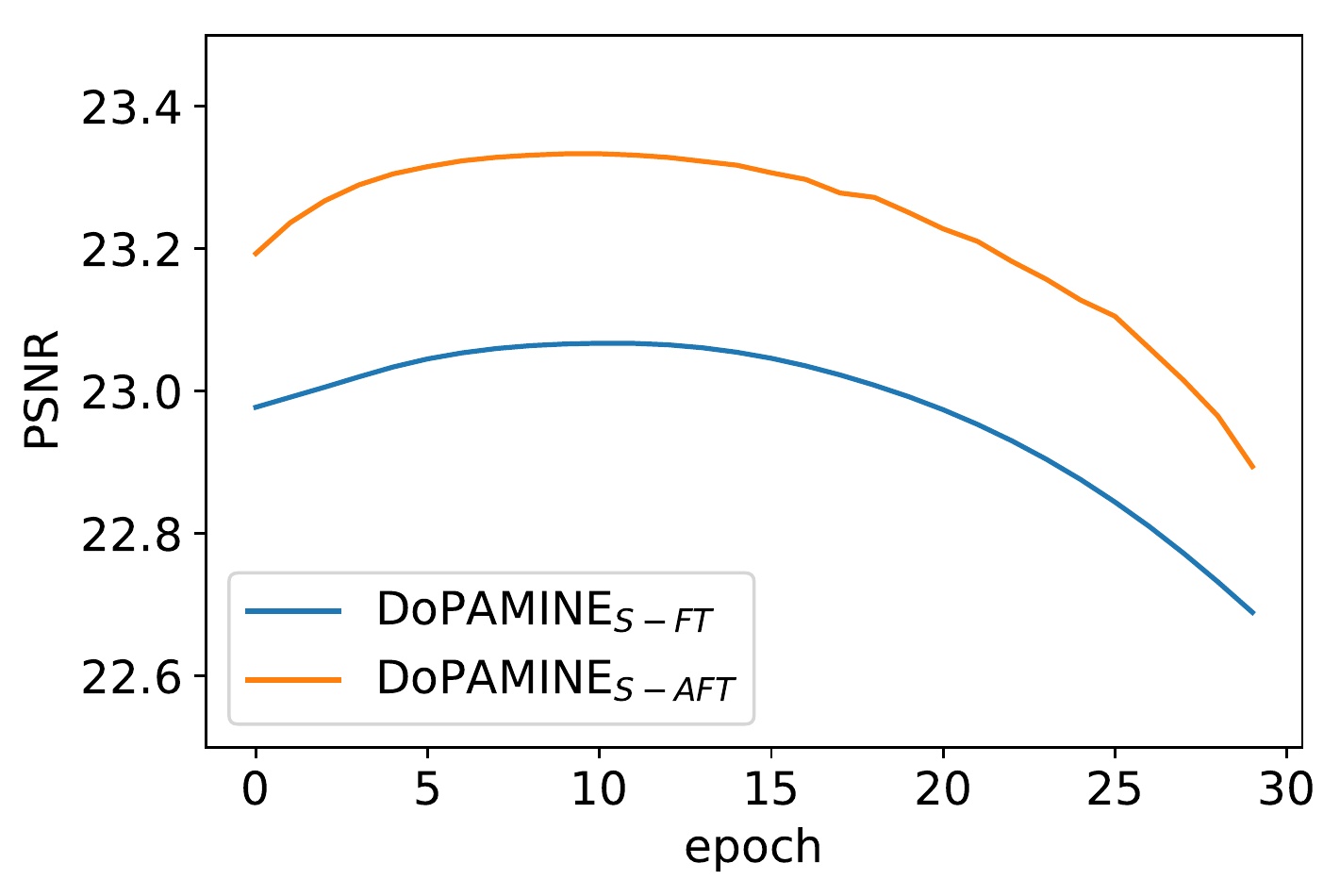}
        \caption{Fine tuning and \textit{AFT}.}
    \end{subfigure}
    \caption{Results of Ablation study. The experiment setting is the same as benchmark experiment with $L=1.0$.}
    \label{fig:ablation}
\end{figure*}

\subsubsection{Results}
Table \ref{tab:PSNR} is the PSNR and SSIM results on the UCML test dataset. SAR-DRN \cite{zhang2018learning} is a state-of-the-art baseline model in despeckling that predicts residual $\Z-\x$ before doing despeckling. It is constructed by 7 dilated convolution layers and skip connections. We reproduced this model and compared with DoPAMINE. The subscript $_S$, $_B$, and $_{-AFT}$ represent supervised, blind, and the data augmented fine tuning ($AFT$), respectively.  

From the table, we first note that the supervised models of DoPAMINE, DoPAMINE$_S$ and DoPAMINE$_B$, always outperforms SAR-DRN$_S$ and SAR-DRN$_B$, respectively, for all $L$, which shows the superiority of our network architecture for the supervised training. Second, we note DoPAMINE$_{S-AFT}$ achieved another gain of 0.3$\sim$0.4$dB$ over DoPAMINE$_S$, which shows the effectiveness and adaptivity of the fine-tuning step. Third, while the performance of DoPAMINE$_B$ is worse than that of DoPAMINE$_S$, we note the AFT of those blind models makes the final model DoPAMINE$_{B-AFT}$ perform almost as well as DoPAMINE$_{S-AFT}$. The implication of this result is the following: in order to attain high denoising performance for various noise levels, we just need to maintain a \emph{single} Blind supervised model as long as fine-tuning with the correct noise variance is possible. Note this approach is much more efficient and powerful than keeping separate supervised model for each noise level or just maintaining a blind model that cannot adapt to specific noise models, like SAR-DRN$_B$.

\subsection{Real SAR image}
In addition to the experiments on the benchmark dataset, we also despeckled real-world SAR-obtained images with 4 number of looks: Death Valley, Flevoland, and San Francisco Bay. These are also widely used benchmark images for SAR despeckling \cite{zhang2018learning}. We only report the result for Flevoland image with $512\times512$ resolution due to space limit. 

In the real SAR images, the underlying clean $\textbf{x}$ is not available, hence PSNR or SSIM cannot be computed. Instead, Equivalent Number of Look (ENL) is one of the widely used metrics for the real SAR image despeckling \cite{argenti2013tutorial}. ENL is defined as 
\begin{align}
    ENL = \frac{\mathbb{E}[\mathbf{\hat{X}}]^2}{\text{Var}[\hat{\mathbf{X}}]}\label{eq:enl}
\end{align}
and is typically evaluated in homogeneous areas. Note in (\ref{eq:enl}), the variance term is in the denominator, hence, the higher the ENL, the more homogeneous the area is. 
However, (\ref{eq:enl}) is not an absolute metric since despeckling models typically have trade-offs between homogeneity and sharpness, namely, models with high ENL often tend to generate blurred images. Therefore, it is also important to visually verify the despeckled images.

\begin{figure*}[ht]
    \centering
    \makebox[20pt]{\raisebox{50pt}{\rotatebox[origin=c]{0}{$\textbf{a}$}}}%
    \begin{subfigure}[b]{0.11\textwidth}
        \includegraphics[width=\textwidth]{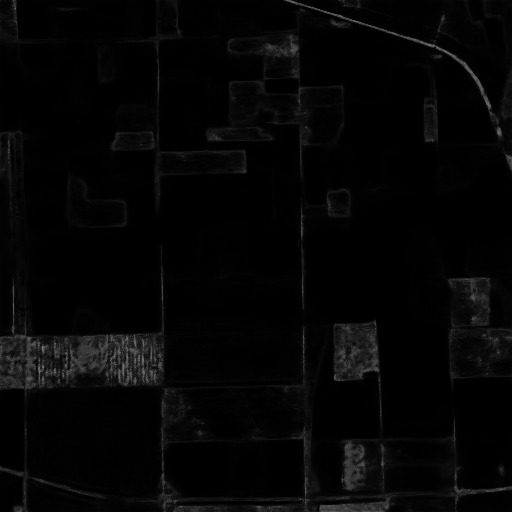}
        \caption{}
    \end{subfigure}
    \begin{subfigure}[b]{0.11\textwidth}
        \includegraphics[width=\textwidth]{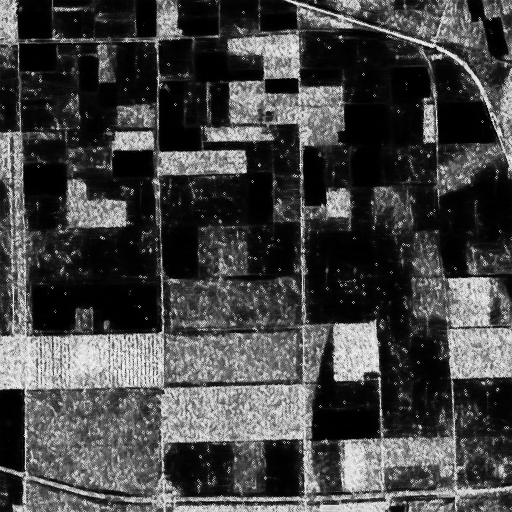}
        \caption{}
    \end{subfigure}
    \begin{subfigure}[b]{0.11\textwidth}
        \includegraphics[width=\textwidth]{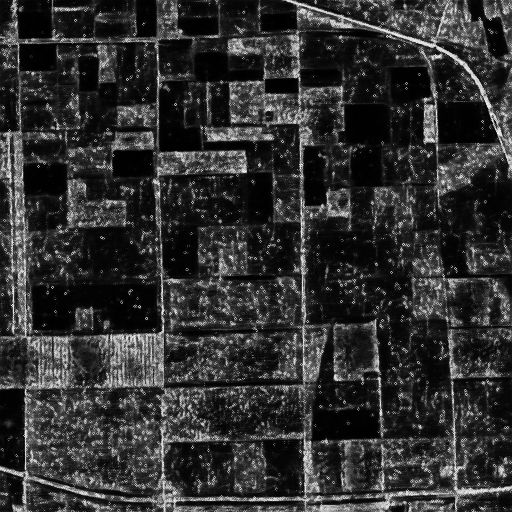}
        \caption{}
    \end{subfigure}
    \begin{subfigure}[b]{0.11\textwidth}
        \includegraphics[width=\textwidth]{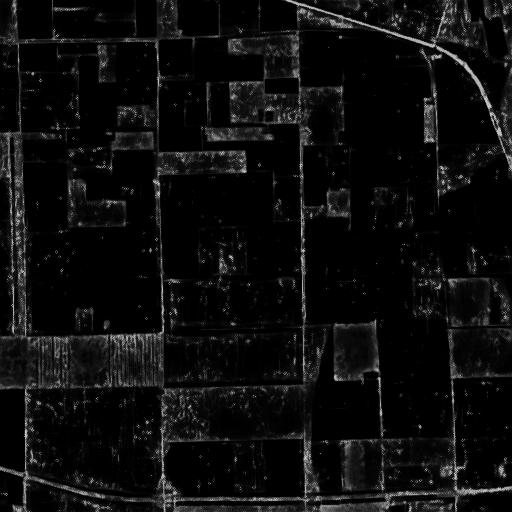}
        \caption{}
    \end{subfigure}
    \makebox[20pt]{\raisebox{50pt}{\rotatebox[origin=c]{0}{$\textbf{b}$}}}%
    \begin{subfigure}[b]{0.11\textwidth}
        \includegraphics[width=\textwidth]{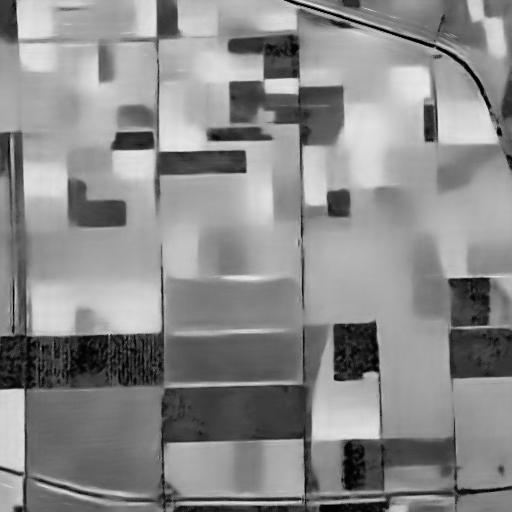}
        \caption{}
    \end{subfigure}
    \begin{subfigure}[b]{0.11\textwidth}
        \includegraphics[width=\textwidth]{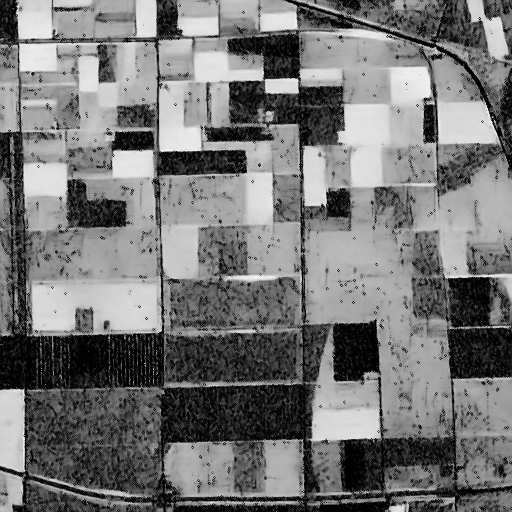}
        \caption{}
    \end{subfigure}
    \begin{subfigure}[b]{0.11\textwidth}
        \includegraphics[width=\textwidth]{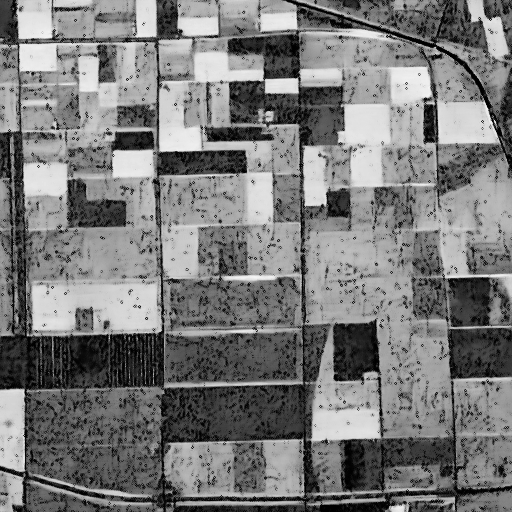}
        \caption{}
    \end{subfigure}
    \begin{subfigure}[b]{0.11\textwidth}
        \includegraphics[width=\textwidth]{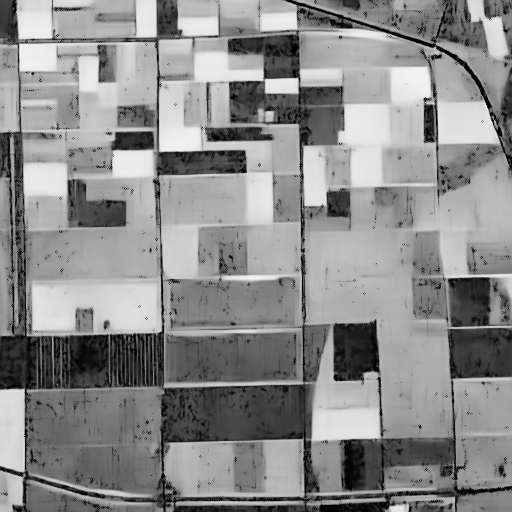}
        \caption{}
    \end{subfigure}
    \caption{Visualization of $\textbf{a}$ and $\textbf{b}$ of DoPAMINE for Flevoland image. From left to right, DoPAMINE$_S$ trained by $L=1.0$, DoPAMINE$_S$ trained by $L=8.0$, DoPAMINE$_B$, and DoPAMINE$_{B-AFT}$ with $L=0.5$ for fine-tuning. }
    \label{fig:AandB}
\end{figure*}

\subsubsection{Results}
Figure \ref{fig:realSar} and Table \ref{tab:ENL} are the results of despeckled Flevoland. 
In Table \ref{tab:ENL}, ENL 1 to 8 correspond to the ENL values on the regions marked with red boxes in Figure \ref{fig:realSar}(a). For supervised models, we used the models with $L=4$ for both SAR-DRN$_S$ and DoPAMINE$_S$, since the number of looks for the image was 4. For DoPAMINE$_{B-AFT}$, we set $L=0.5$, learning rate to $8\times 10^{-6}$, and the number of epochs to 6 for carrying out the fine-tuning. %We found that the performance was not very sensitive to those hyperparameters. 

In Table \ref{tab:ENL}, we observe that DoPAMINE$_{B-AFT}$ achieves the highest average ENL values compared to others. Note the ENL values for  DoPAMINE$_{B}$ is very low, but the AFT step improves the metric significantly and surpasses that of SAR-DRN$_S$. Figure \ref{fig:realSar} visualizes the despeckling results of comparing methods in Table \ref{tab:ENL}. We first observe that SAR-DRN$_S$ is more blurred than other images, especially left above region, which shows that high ENL values do not always translate to good despeckling quality in all image regions. In contrast, we can see that DoPAMINE$_B$ generates much sharper images than SAR-DRN$_S$ or DoPAMINE$_S$. However, it contains black spots in both dark and bright regions, while other models tend to contain black spots only in the dark regions. When checking the despeckling results on the benchmark dataset, we noticed that black spots typically occurs when $\sigma$ of the training set is smaller than that of the test image. Motivated by this, for DoPAMINE$_{B-AFT}$, we used smaller $L$ (\emph{i.e.}, larger $\sigma$) in the AFT loss (\ref{eq:aft_loss}), and the resulting image is given in Figure \ref{fig:realSar}(e). We can see from the figure that most of the black spots are now removed, while preserving the details and sharpness of the image. The magnified red box shows the difference of the despeckling more closely. We can clearly observe the quality of our DoPAMINE$_{B-AFT}$ is superior to other methods, showing the power of our method that can adaptively calibrate the image qualities by choosing appropriate $L$. Note that this is a unique property of DoPAMINE, as other deep learning based method cannot adapt to the noisy image.

\subsection{Ablation study}

\subsubsection{Scale add layer}
To validate the effectiveness of the scale add layer (SA), we carried out two experiments. In Figure \ref{fig:ablation}(a), we tested with widely used ResNet20 architecture to show the generality of our SA. The figure represents initial variance of the feature maps after each convolution layer of ResNet20 models, when the input to the network was standard zero mean Gaussian and the weights were initialized with the He initialization \cite{he2016deep}. Note ordinary ResNet20 architecture has many addition layers due to the skip connections. ResNet20-SA is the model with the same architecture, but with addition layer replaced with SA. Vanilla CNN is the model that does not have skip connections, hence it has no addition layers. From the figure, we clearly observe that He initialization alone is not enough to control the variances of the feature maps when there are many addition layers (or skip connections) in the architecture, since the variance of ResNet20 keeps exponentially increasing as the layer increases. Note for Vanilla CNN, the variance is kept to be constant, as expected with the He initialization. In contrast, we note ResNet20-SA still maintains the constant variance even though there are many skip-connections, since the SA re-scales the feature map so that the variance is controlled. This property suggests that the training of the network with SA may be accelerated compared to the one without it. Figure \ref{fig:ablation}(b) confirms such result and shows that our DoPAMINE$_S$ with SA is more stable and achieves high PSNR faster than the one with ordinary addition layer with batch normalization. In this case, when only the addition layer was used, the PSNR was too low to show in the same figure.

\subsubsection{ResNet blocks and training choices}
Figure \ref{fig:ablation}(c) shows the PSNR results of DoPAMINE$_S$ model with and without the final ResNet block shown in Figure \ref{fig:DoPAMINE}. We note the model with the ResNet block outperforms the one without it, which justifies our choice of the architecture. 

Figure \ref{fig:ablation}(d) compares the models with and without the $a_i$ term in output of the network. Namely, the model without $a_i$ simply reconstructs $x_i$ only with the bias term $b_i$, hence, it does not use the noisy pixel $Z_i$. The curves in Figure \ref{fig:ablation}(d) that only have subscript `B' stand for those models. The figure shows both the supervised-only and fine-tuning results and confirms the necessity of using the slope parameter $a_i$ for learning the pixel-wise despeckling mappings.

Figure \ref{fig:ablation}(e) compared the PSNR of $FT$ and $AFT$ on DoPAMINE. We see the maximum PSNR of $AFT$ is 0.3$dB$ higher than $FT$, and this is in line with the findings in \cite{cha2018fully}. This result also show that the performance of $FT$ and $AFT$ is robust to epochs around 5 to 15. 

\subsubsection{Visualization of $\textbf{a}$ and $\textbf{b}$}
Figure \ref{fig:AandB} visualizes the coefficients of the affine mappings, $\textbf{a}$ and $\textbf{b}$, for several variations of DoPAMINE. Note that Figure \ref{fig:AandB}(d),(h) are the $\textbf{a}$ and $\textbf{b}$ of the model for Figure \ref{fig:realSar}(e).
We can see that if
$\textbf{a}$ tends to have low values, $\textbf{b}$ become more homogeneous, but it loses sharpness. In contrast, if $\textbf{a}$ tends to have  high values, $\textbf{b}$ gets sharper, but it loses homogeneity. 
Hence, we can say that the values that $\textbf{a}$ take determines the sharpness or homogeneity of an image.
Next, we describe how $\textbf{a}$ and $\textbf{b}$ are changed during $AFT$. 
According to equation (5), $Z_i$ and $\sigma$ acts as effective regularization constant for $a_i$. We cannot control $Z_i$, however, by changing $\sigma$, we can control the values that $a_i$ takes. As a result, by adjusting $\sigma$, we can determine the amount of sharpness or homogeneity of the despeckled image.

\section{Concluding remarks}

In this paper, we showed that our DoPAMINE framework has an ability to fine-tune and outperforms the state-of-the-art model on both benchmark and real images.
There are several extendable future approaches. First, more complex mapping functions and their unbiased estimators can be derived. Second, we are going to check whether our framework can be applied to other domains, such as medical image despeckling. 

\section*{Acknowledgments}

This work is supported in part by the Basic Science Research
Program through the National Research Foundation of
Korea [NRF-2016R1C1B2012170], by the ICT R\&D program
of MSIT/IITP [2016-0-00563], and by the ITRC support program  of MSIT/IITP [IITP-2018-0-01798].

%\newpage
%\bibliography{7_Citation.bib}

\bibliographystyle{aaai}

\end{document}